i
\documentclass[%
reprint,
superscriptaddress,
amsmath,amssymb,
aps,
twocolumn,
float,
floatfix,
]{revtex4-1}

\usepackage{graphicx}
\usepackage{xcolor}
\usepackage{soul}
\usepackage{dcolumn}
\usepackage{bm}
\usepackage{siunitx}

\newcommand{\fa}{$\langle f_{1} \rangle$} 
\newcommand{\fb}{$\langle f_{2} \rangle$} 
\newcommand{\fc}{$\langle f_{3} \rangle$} 
\newcommand{\Cov}{\mathrm{Cov}}

\setstcolor{red}

\begin{document}
	\title{ Weak-winner phase synchronization: A curious case of weak interactions}
	
	\author{Anshul Choudhary}
	\affiliation{Theoretical Physics/Complex Systems, ICBM, Carl von Ossietzky University of Oldenburg, 26129 Oldenburg, Germany}
	\affiliation{Nonlinear Artificial Intelligence Laboratory, North Carolina State University, USA}
	\author{Arindam Saha}%
	\affiliation{Theoretical Physics/Complex Systems, ICBM, Carl von Ossietzky University of Oldenburg, 26129 Oldenburg, Germany}
	\author{Samuel Krueger}
	\affiliation{Department of Physics, Illinois State University, Normal, IL 61790, USA}
	\author{Christian Finke}
	\affiliation{Theoretical Physics/Complex Systems, ICBM, Carl von Ossietzky University of Oldenburg, 26129 Oldenburg, Germany}
	%
	\author{Epaminondas Rosa, Jr.}
	\affiliation{Department of Physics, Illinois State University, Normal, IL 61790, USA}
	\author{Jan A. Freund}
	\affiliation{Theoretical Physics/Complex Systems, ICBM, Carl von Ossietzky University of Oldenburg, 26129 Oldenburg, Germany}
	
	\author{Ulrike Feudel}
	\affiliation{Theoretical Physics/Complex Systems, ICBM, Carl von Ossietzky University of Oldenburg, 26129 Oldenburg, Germany}
	%
	
	\date{\today}
	
	\begin{abstract}
		We report the observation of a non-trivial emergent state in a chain of non-identical, heterogeneously coupled oscillators 
		where a set of weakly coupled  oscillators becomes phase synchronized while the strongly coupled ones remain drifting. 
		This intriguing ``weak-winner" synchronization phenomenon can be explained 
		by the interplay between non-isochronicity and natural frequency of the oscillator, as coupling strength is varied.
		Further, we present sufficient conditions under which the weak-winner phase synchronization
		can occur for limit cycles as well as chaotic oscillators. Employing a model system from ecology as well as a paradigmatic model from physics, we demonstrate that this phenomenon is a generic feature for 
		a large class of coupled oscillator systems. The realization of this peculiar, yet quite generic weak-winner dynamics, can have far reaching consequences in a wide range of scientific disciplines that deal with the phenomenon of phase synchronization, including synchronization of networks. Our results also highlight the role of non-isochronicity~(shear) as a fundamental feature of an oscillator in shaping emergent dynamical patterns in complex networks.
		
	\end{abstract}
	\maketitle
	
	\section{\label{sec:level1} Introduction}
	Interactions play a fundamental role in nature since many functions, for instance, sensory or information processing rely on collective tasks, involving an exchange of matter or energy, rather than on individual entities.
	One of the oldest examples of such collective behavior has originated from the physics of coupled pendulums, which are able to \emph{synchronize} their motion in time through a weak mechanical coupling~\cite{huygens}. Since its discovery, synchronization has been observed and studied in many areas of science with problems ranging from collective behavior of a large population of chemical oscillators~\cite{sync_chemical} as well as spiking and bursting of neurons in neural networks~\cite{sync_neural1,sync_neural2} to coupled superconducting Josephson arrays~\cite{sync_josephson} and  information transfer in neural systems~\cite{sync_information}, among others (\cite{strogatz_sync} and references therein).  
	Mutual synchronization implies the emergence of \emph{coherence} in the system through the adjustment of internal rhythms of individual entities without the presence of any central point of control. Several interesting classifications of this broad phenomenon have emerged through extensive research done in the last few decades, namely complete synchronization~(CS)~\cite{sync_complete}, generalized synchronization~(GS)~\cite{sync_generalized} and phase synchronization~(PS)~\cite{phase_sync_prl}. CS implies that the coupled systems remain in step with each other for all times after transients. However, CS can only occur in a system of coupled identical units. By contrast, GS is a state where the coupled elements maintain a functional relationship with each other for all times after transients. Note that GS can be realized for systems where non-identical units are coupled.
	In this study our focus is on the phenomenon of \emph{phase synchronization}~(PS) in coupled systems which is characterized by oscillators keeping their phases in step with each other while showing no correlation between their amplitudes~\cite{phase_sync_prl}. It is  one of the most ubiquitous phenomena in coupled oscillator systems, pervading both the natural and technological world~(\cite{pikovsky_sync}  and references therein).    
	One of the central problems concerning PS is to explain the mechanism(s) behind its emergence for different dynamical behaviors such as limit cycle oscillations, quasi-periodicity and chaos and also for different coupling topologies such as ring, star and small-world networks~\cite{network_topology}. The contemporary approach essentially relies on the fact that PS emerges out of the complex interplay between coupling and frequency detuning~\cite{kuramoto_rmp,ps_laser,PS_classic}. 
	\begin{figure*}[ht]
		\centering
		\includegraphics[width=0.98\textwidth]{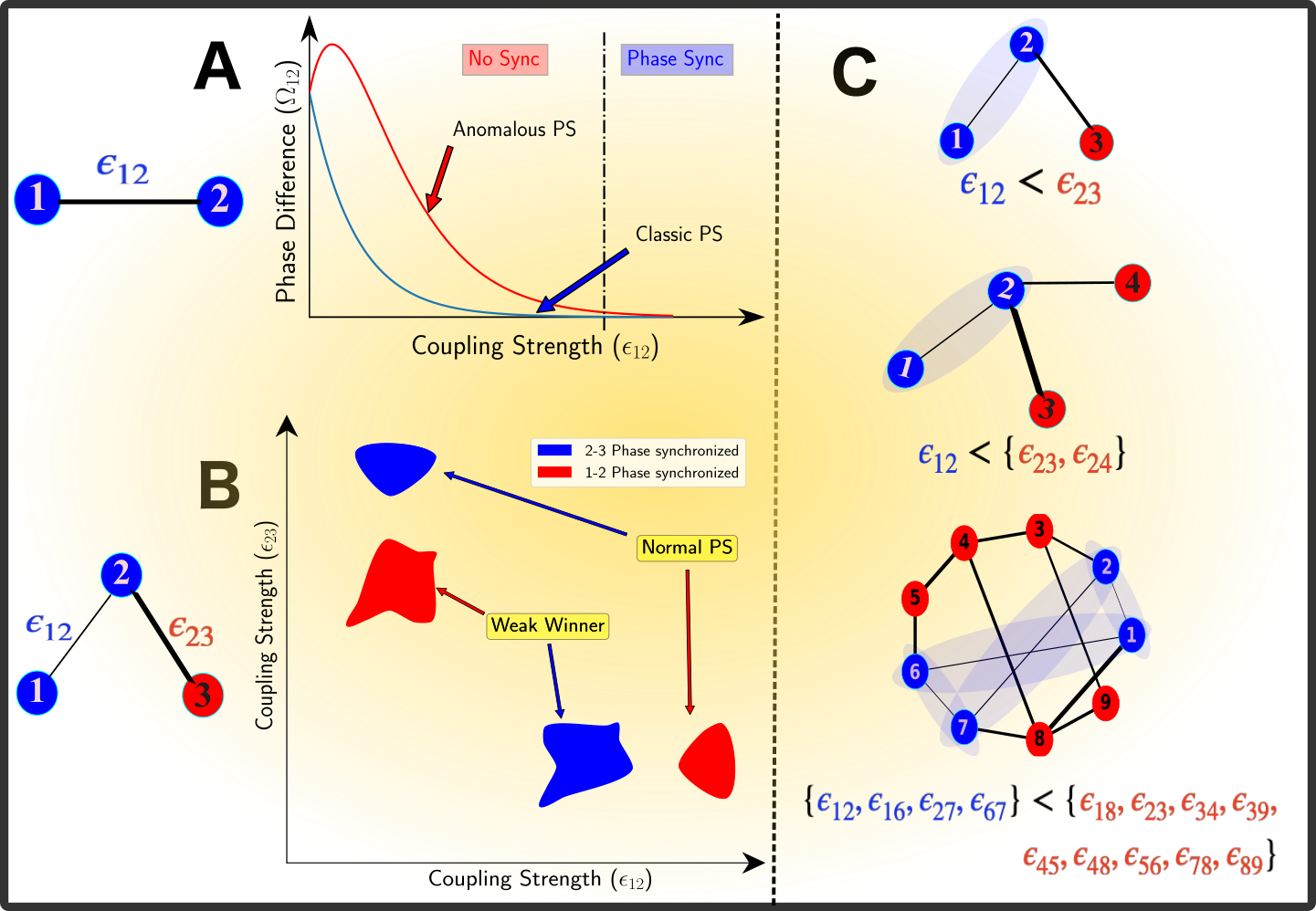}
		\caption{ A brief summary of the essence of the weak-winner phenomenon and how it can manifest in progressively larger networks. Panel {\bf (A)} depicts the two distinct routes to phase synchronization in a pair of coupled oscillators. One is the classic monotonic decay (blue curve) in phase difference and the other is anomalous phase synchronization with non-monotonic decay (red curve) of phase difference[16-20, 36] as coupling is increased. 
			Panel {\bf (B)} shows how ``weak-winner" emerges in a chain of three heterogeneously coupled oscillators. For three coupled oscillators, there are at least two links and one of them could be a weaker one, the weakly linked pair of oscillators synchronize their phases while the other pair with stronger coupling strength remains drifting (details discussed in Section III). This phenomenon where a subset of a network with weakly linked nodes synchronizes their phases while the rest of the network with strongly linked nodes remains drifting is termed as ``Weak-Winner".
			Panel {\bf (C)} shows some examples of weak-winner phase synchronization as the network size is increased. The coupling strength between oscillators are reflected by the edge thickness. The oscillators colored cyan are in phase synchrony.    }
		\label{fig:WW_summary}
	\end{figure*}
	~However, in this study we present an intriguing type of PS which cannot be explained by the aforementioned approach. This state which we call ``weak-winner" is an emergent dynamical pattern in heterogeneously coupled oscillators chain where the weakly linked part of the chain exhibits phase synchrony while the strongly couple part remains incoherent. Further, we suggest a mechanism utilizing the concept of non-isochronicity~\cite{anomalous_pre,anomalous_experiment,anomalous_exp2, anomalous_noise,anomalous_timeseries,isochron_prx} to explain the emergence of this non-trivial state of PS.
	
	Formally, two coupled
	oscillators can be considered phase synchronized
	if $\Delta\varphi = |\varphi_{1} - \varphi_{2} | < \text{constant}$ for sufficiently long periods of time. Here $\varphi_{1}$ and $\varphi_{2}$ are the
	phases of the two oscillators and the constant, for the purpose of our study, is say $2\pi$. Conversely, phase synchrony
	breaks down whenever one of the oscillators advances its
	phase at least a full $2\pi$ cycle ahead of the other~\cite{rosa_prl}.
	In general, increasing the coupling strength between several oscillators synchronizes their phases. Nonetheless, we show here that surprisingly the phase synchronization can also appear in the weakly coupled part of the network while the strongly coupled part remain desynchronized (see Fig.~\ref{fig:WW_summary}).

\begin{figure*}[htb]
		\centering
		\includegraphics[scale=0.50]{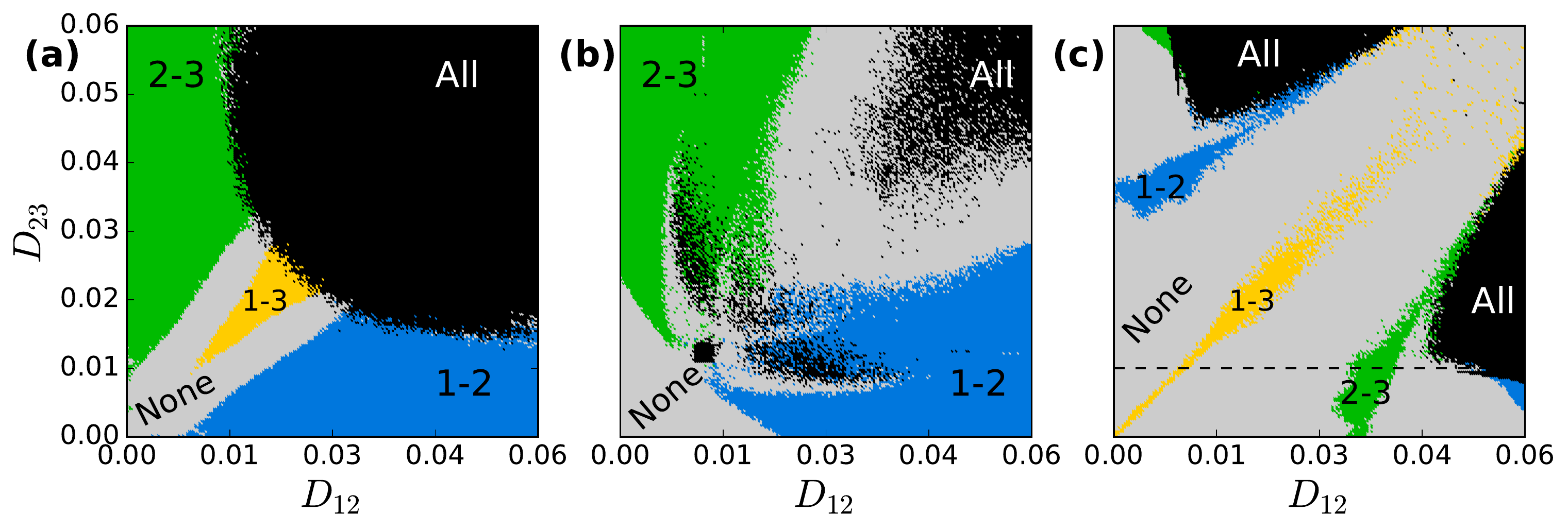}
		\caption{  Parameter space plots showing, as indicated by the labels(and colors), regions of $D_{12}$ and 
			$D_{23}$ values for oscillator pairs in phase synchrony or not, with  (a) $b_{2}=1.00$, 
			(b) $b_{2}=1.03$ and (c) $b_{2}=1.1$.}
		\label{fig:fig1}
	\end{figure*}

	We first demonstrate this using a minimalistic setup with three coupled oscillators. One of the oscillators (say oscillator 2) is coupled
	bi-directionally to the two other oscillators (say oscillators 1
	and 3) with coupling parameters $D_{12}$ and $D_{23}$ respectively and
	there is no direct coupling between oscillators 1 and
	3 (see figure~\ref{fig:fig1}). This linear chain setup has been studied, for example, 
	in the context of R\"ossler systems~\cite{breban_ott_pre} and chaotic
	lasers~\cite{mcallister_pre} forced by two sinusoidal signals, three
	coupled R\"ossler systems exhibiting partial phase synchronization~\cite{schelter_prl} 
	and competing synchronization~\cite{nishikawa_physicaD}, and three
	coupled semiconductor lasers as well as three neurons displaying 
	relay synchronization~\cite{fischer_prl}.

	We observe both competing and relay PS~\cite{relaySync1,relaySync2}
	in our three oscillators system and in-addition a counter-intuitive type
	of phase synchronization. The latter happens for certain regions in the
	$D_{12}-D_{23}$ parameter space where the two weakly coupled
	oscillators do stay in phase synchrony with each other while the
	two strongly coupled oscillators do not. 
	This unexpected behavior, which we call ``weak-winner phase synchronization" can be understood 
	as a result of the complex interplay between 
	shear (non-isochronicity) and the natural frequency of individual oscillators 
	as the coupling strength is varied.
	We also present sufficient conditions under which 
	a coupled oscillator system can exhibit weak-winner PS. Further,
	we claim that this phenomenon is a generic feature of a large class of coupled nonlinear 
	oscillator systems and provide examples which validate our claim.
	\section{\label{sec:level2}Model and First Observations}
	To demonstrate the variety of possible applications of weak-winner PS, we first use an example from theoretical ecology to discuss the case of chaotic PS.\medskip\\
	\emph{Chaotic oscillator model.---} We consider three coupled chaotic oscillators $\left( i=1,2,3 \right)$, each of which represents a
	food chain with three trophic levels at a particular spatial location (patch). This model was originally developed
	to demonstrate phase synchronization in population dynamics~\cite{blasius_nature}. 
	Each of the three population patches consists of nutrients (vegetation) $x_{i}$, prey (herbivore)
	$y_{i}$ and predator (carnivore) $z_{i}$ as species. The coupling between the patches accounts for possible migration of herbivores and carnivores.
	Hence, the dynamics of the entire system is given as
	\begin{subequations}\label{eq:SystemEqns}
		\begin{align}
			\dot{x}_{i} &= a_{i}x_{i} - \frac{\epsilon_{1}x_{i}y_{i} }{(1+k_{1}x_{i} )}\\
			\dot{y}_{i} &= -b_{i}y_{i} + \frac{\epsilon_{1}x_{i}y_{i} }{(1+k_{1}x_{i} )} -\epsilon_{2}y_{i}z_{i} +\sum_{j=1}^{3}D_{ij}(y_{j} -y_{i} )\\
			\dot{z}_{i} &= -c_{i}(z_{i}-\zeta_{i}) + \epsilon_{2}y_{i}z_{i} +\sum_{j=1}^{3}D_{ij}(z_{j} -z_{i} ) 
		\end{align}
	\end{subequations}
	
	where $a_{i}$ represents the vegetation 
	growth rate, and $b_{i}$ and $c_{i}$ represent, respectively,
	the herbivore and carnivore mortality rates in the absence of interspecies 
	interactions. The terms $\frac{\epsilon_{1}x_{i}y_{i} }{(1+k_{1}x_{i} )}$ denoting 
	vegetation-herbivore interaction
	(prey growth rate), and $\epsilon_{2}y_{i}z_{i}$ describing herbivore-predator 
	interaction, are the standard Holling Type II
	and Lotka-Volterra functions, respectively. The parameter $\zeta_{i}$  
	accounts for the availability of food for the predator in addition 
	to its preferred prey~\cite{ecology_book}. Parameters
	$D_{ij} = D_{ji}$ represent the coupling strength between patch {\it i} and {\it j} representing the migration of herbivores and carnivores between the patches. For this study, we assume that the three patches are connected in a linear chain, which results in a coupling matrix ${\bf D} = \Bigl( \begin{smallmatrix}
	0 & D_{12} & 0 \\
	D_{12} & 0 & D_{23} \\
	0 & D_{23} & 0 
	\end{smallmatrix} \Bigr)$. 
	We fix the parameters at
	$a_{1} = a_{2} = a_{3} = 1.0$, $b_{1} = b_{2} = b_{3} = 1.0$,
	$c_{1} = c_{2} = c_{3} = 10.0$, $\epsilon_{1} = 0.25, \epsilon_{2} = 1.0$, $k_{1} = 0.05$, and
	$\zeta_{1}=\zeta_{2}=\zeta_{3}=0.006$, unless specified otherwise. For this parameter set, the population densities exhibit
	chaotic oscillations which resemble those of the R\"ossler
	system~\cite{rossler_equation} with phase coherent dynamics. 
	
	This means that the trajectory oscillates chaotically around a fixed center of rotation, and on a two-dimensional projection of the attractor, an instantaneous phase
	can be defined as the increasing angle between an arbitrarily 
	fixed reference axis and the radius of the trajectory~\footnote{For the range of parameter values used in this work, the individual oscillators of our system remain phase coherent. However, the methods we employ for detecting and analysing phase synchronization are applicable to phase non-coherent oscillators as well, as long as the adequate techiques are used for obtaining their phases. See, for example, R. Follmann, E. Rosa, Jr. and E. E. N. Macau, Phys. Rev. E, 83, 016209 (2011).}.  
	
	All numerical simulations presented here were performed with the
	Dormand-Prince (DOPRI5) adaptive step size algorithm~\cite{dopri5}. 
	To detect 1$\colon$1 phase synchrony between oscillator $i$ and $j$,
	we compute their unwrapped instantaneous phases $\varphi_{i}(t)$ and $\varphi_{j}(t)$ and check for,
	\begin{align}
		\delta\varphi_{ij} = std\left(|\varphi_{i}(t) - \varphi_{j}(t) |\right) < 2\pi, ~~~ \forall t > t_{trans}
	\end{align}
	where, std(.) is the standard deviation
	and the transient time, $t_{trans}$ is taken to be $10^6$ arbitrary time units.

	To study how the coupling strengths (migration rates) affect the phase dynamics 
	among the three oscillators, we generate plots in coupling
	parameter space indicating different synchronous behaviors (Fig.~\ref{fig:fig1}). The values of $D_{12}$ and $D_{23}$ vary in the
	range between 0.00 and 0.06. We keep $b_{1} = b_{3} = 1.00$ in all
	three plots, and use $b_{2} = 1.00$, $b_{2} = 1.03$ and $b_{2}  = 1.1$
	in plots (a), (b) and (c), respectively, indicating that environmental conditions for the herbivores are identical in the outer two patches, but differ in the central one. In fact, a small increase
	in the prey mortality parameter $b_{i}$ causes a slight increase
	in the intrinsic frequency of the $i^{th}$ oscillator.
	Figure~\ref{fig:fig1}(a) representing the case of three
	coupled identical oscillators conspicuously displays five
	different parameter regions characterized by different states of phase synchrony
	among oscillators: (i) synchronous behavior between oscillator 1 and 2 only, labeled 1-2 (blue in color),
	(ii) synchronous behavior between 2 and 3 only, labeled 2-3 (green in color), (iii) no synchronization between any pair of oscillators labeled None (gray in color), (iv) relay synchronization between the two outer oscillators 1 and 3, labeled 1-3 (yellow in color), and (v) complete synchronization of all three oscillators, 
	labeled All (black in color).
	

	The size and location of the synchronization regions
	change when we increase the $b_{2}$ value to $1.03$ (Fig.~\ref{fig:fig1}(b)). We now see that the None-synchronized region is enlarged at the expense of complete synchronization, while the
	1-2 and 2-3 synchronized regions are not significantly affected. 
	The original phase structure (Fig.~\ref{fig:fig1}(a)) 
	gets some distortion while still maintaining its symmetry.
	Note that relay synchronization disappears completely in this case.
	As we advance the $b_{2}$ value further to $1.1$ (Fig.~\ref{fig:fig1}(c)), all five different parameter regions 
	found in Fig.~\ref{fig:fig1}(a) are also present, with two regions
	of particular interest. Notice on the upper-left quadrant
	the 1-2 synchronized (blue	 in color) region for weak $D_{12}$
	coupling and strong $D_{23}$ coupling. Due to the stronger
	$D_{23}$ coupling, this parameter region would be expected
	to generate 2-3 phase synchronization, not 1-2 as it does. Analogously, 
	due to the symmetry in our setup we find a 2-3 phase
	synchronization region
	, with strong $D_{12}$ and weak $D_{23}$ coupling. We call this phenomenon a weak-winner phase synchronization. 
	
\begin{figure}[hbp]
		\centering
		\includegraphics[width=0.4\textwidth,height=0.3\textheight]{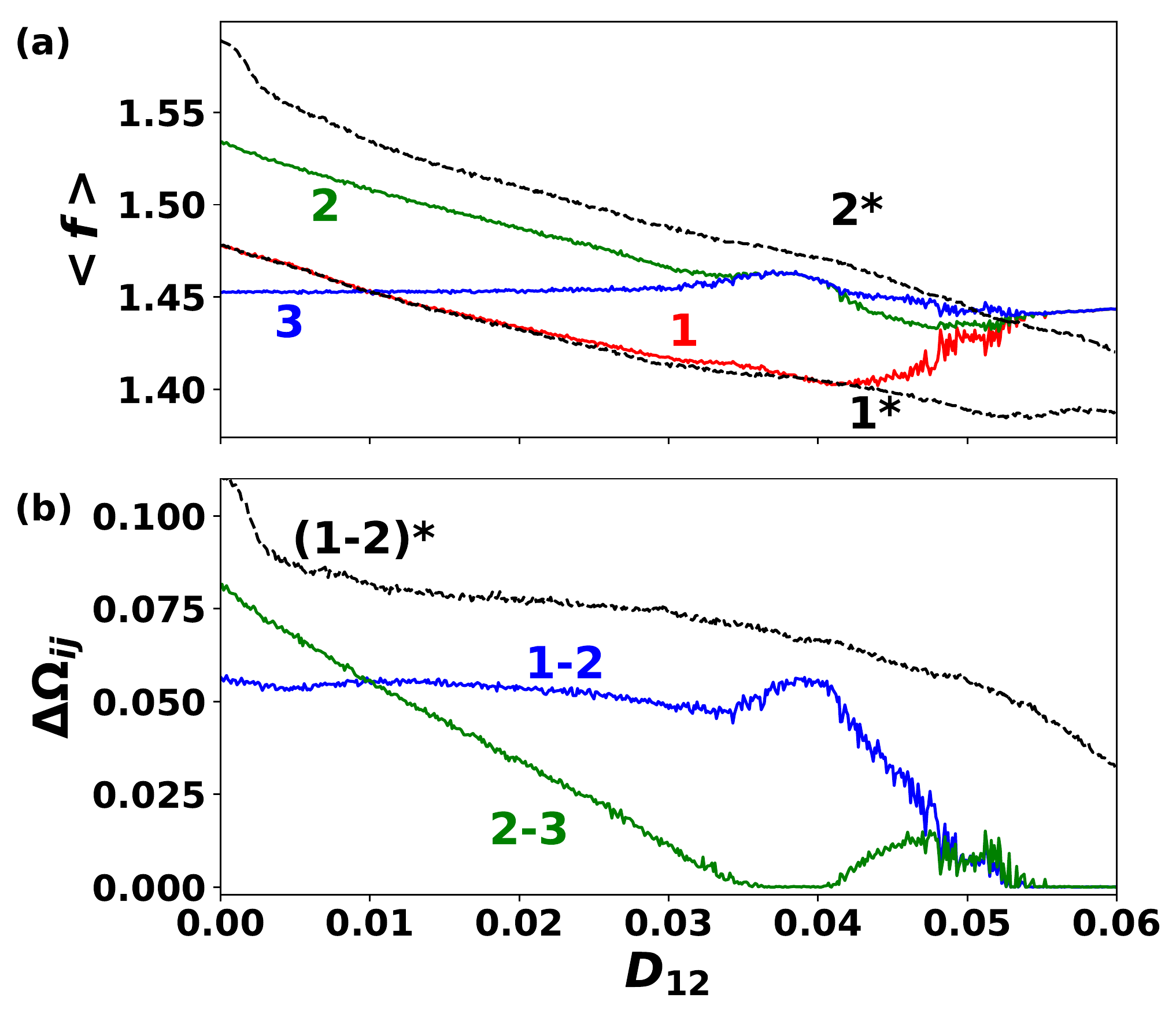}
		\caption{ \small
			(a) Variation of mean frequency with coupling strength $D_{12}$ for $b_{2}=1.1$. Corresponding to (a), (b) represents the variation of mean relative frequency for oscillator pairs 1-2 and 2-3, as labeled, with $D_{12}$ for a fixed value of $D_{23}=0.01$(along horizontal line in the Fig.~\ref{fig:fig1}(c)). For comparison, lines $1*$ and $2*$ correspond to the case when $D_{23}=0$}
		\label{fig:fig2}
	\end{figure}

	%
	This seemingly counterintuitive PS, 
	where the weak coupling wins over the strong coupling for 
	synchrony, can be corroborated by observing how
	the mean frequencies of the oscillators \fa, 
	\fb, \fc ~and their mutual differences: $\Delta\Omega_{ij} = | \langle f_{i} \rangle - \langle f_{j} \rangle |$, vary with changes in the coupling strength. 
	To be specific, we fix $D_{23}=0.01$ and vary $D_{12}$ in the interval
	[0, 0.06], as indicated by
	the horizontal line in Fig.~\ref{fig:fig1}(c). The mean frequencies
	are depicted in Fig.~\ref{fig:fig2}(a) by the lines labeled 1,
	2 and 3 respectively, corresponding to the case exhibiting weak-winner PS (Fig.~\ref{fig:fig1}(c)).
	Initially separated and distinct, the lines evolve
	for increasing values of $D_{12}$, showing a tendency for \fa
	~and \fb ~to decrease and for \fc ~to remain about constant,
	corresponding to the
	case of weak and constant $D_{23} = 0.01$ with growing
	$D_{12}$. 
	When \fb ~and \fc ~become equal, weak winner phase synchrony appears.
	Interestingly, systems 1 and 
	2 synchronize more easily when system 3 is coupled to
	system 2, as opposed to the case when $D_{23} = 0$ denoted
	by lines $1^{*}$ and $2^{*}$. In fact, system 3 works as a catalyst,
	causing systems 1 and 2 to synchronize earlier, i.e. for smaller
	$D_{12}$ values compared to the case when system 3 is not
	part of the process.  
	
	So far, we have observed that due to some interplay between coupling and 
	frequency mismatch, one could get a very unexpected synchronized state \textendash ~ the weak-winner phase synchronization. At first sight, the emergence of weak-winner PS might appear to be the consequence of a phenomenon known as short wavelength bifurcation~(SWB)~\cite{Sync_MSF}. However, this is not the case as explained in Appendix D.
	Instead, we can explain the mechanism of the emergence of such synchronized state as a result of the existence of anomalous phase synchronization(APS)~\cite{anomalous_pre,anomalous_experiment,anomalous_exp2, anomalous_noise,anomalous_timeseries, pikowsky_aps, APS_new1}. To demonstrate this in detail we recall briefly the concept of APS. 
	For a system of two coupled oscillators, APS is a state whereupon the frequency difference between the oscillators shows a non-monotonic behavior with respect to the coupling strength. Instead of monotonically decreasing, the frequency difference increases for certain coupling range before its inevitable decay with increasing coupling strength (see Appendix C for the intuitive understanding behind APS). 
	This non-monotonic relationship between coupling and the spread of frequencies occurs when $ Cov(\omega_{i},q_{i})>0$, where 
	\begin{equation}
	Cov(\omega_{i},q_{i})  =   (\omega_{i}-<\omega_{i}>)(q_{i}-<q_{i}>) 
	\end{equation}
	$\omega_{i}$ and $q_{i}$ are the natural frequency and shear (non-isochronicity) of the $i^{th}$ oscillator~\cite{anomalous_pre}. Now, for our system (Eq.~\ref{eq:SystemEqns}), we do see the
	signatures of APS as shown in Fig.~\ref{fig:fig2}(b), where the frequency difference of oscillator pair 1-2 varies non-monotonically with coupling.
	However, to fully analyze the system, one must have a clear definition of shear in the system. Generally, both shear ($q$) and
	natural frequency ($\omega$) are functions of the system parameters 
	and in order to check the $\Cov[\omega,q]$, these functions need to be determined. While it is possible to approximate these functions numerically for any nonlinear oscillator, we find it more convincing to study a paradigmatic system which has both shear and natural frequency explicitly present in the governing equations as system parameters.
	\section{\label{sec:level3} Mechanism: A Paradigmatic Model approach }
	To explore the mechanism of the emergence of weak-winner PS, we turn to a simpler model which is known to exhibit APS and which also contains frequency and shear as system parameters.\medskip\\
	\emph{Limit cycle model.---} Here, we are going to use the same
	coupling structure as before but with individual
	oscillators represented by complex Stuart-Landau equations. The Stuart-Landau equation represents a generic mathematical equation describing the behavior of any nonlinear oscillator close to the onset of oscillations. Therefore, in this system the oscillators exhibit only limit cycles when uncoupled and no chaotic oscillations. Interestingly, the extension of the Stuart-Landau equation to spatial domains is given by the complex Ginzburg-Landau equation which is one of the most widely studied nonlinear equation in the physics community describing a plethora of phenomena ranging from superconductivity~\cite{superconductivity_GL} and Bose-Einstein condensation~\cite{BEC_GL} to nonlinear waves and chemical oscillations~\cite{chemical_oscillations_GL}.

	The governing dynamics of the Stuart-Landau system is determined by:
	\begin{equation}\label{eqn:SL_eqns}
	\dot{z}_{j} = z_{j}\left[1+i(\omega_{j}+q_{j}) -(1+iq_{j})|z_{j}|^2  \right] \\
	+\sum_{k=1}^{3}D_{jk}(z_{k} -z_{j} )
	\end{equation}
	\medskip
	where $z_{j} = \rho_{j}e^{i\theta_{j}}$ and $j=1,2,3$.  Here, $\omega_{j}$ represents
	the intrinsic frequency of the oscillator j and $q_{j}$ is the degree of non-isochronicity (or shear)
	which is basically a measure of the dependence of the frequency on the amplitude of the oscillator. In this model, shear and frequency are system parameters. 
	In order to understand the phase dynamics of the system,
	we reduce Eq.~(\ref{eqn:SL_eqns}) to a pure phase equation that is valid in the weak coupling limit, given by :	 
	\begin{equation}
	\dot{\theta}_{j} = \omega_{j} + q_{j} \sum_{k=1}^{3} D_{jk}  + \sum_{k=1}^{3} D_{jk} \left[ \sin \phi_{jk} - q_{j} \cos \phi_{jk} \right],
	\label{eq:Phase_1}
	\end{equation}
	where $\phi_{mn} = \theta_{n} - \theta_{m}$ is the relative phase between oscillator $m$ and $n$. Equation (\ref{eq:Phase_1}) can be further represented in terms of the evolution of relative phases as:
	\begin{align}
		\dot{\phi}_{12} &= C_{1} - A_{1} \sin \left( \phi_{12} + \alpha \right) - B_{1} \sin \left( \phi_{32} + \beta \right) \label{eq:Phase_Diff_Final_1} \\
		\dot{\phi}_{32} &= C_{2} - A_{2} \sin \left( \phi_{12} + \beta \right) - B_{2} \sin \left( \phi_{32} + \alpha \right)
		\label{eq:Phase_Diff_Final_2}
	\end{align}
	
	\medskip
	where the constants, $C_{1} = \Delta \omega + D_{12} \Delta q + D_{23} q_{2}$, $C_{2} = \Delta \omega + D_{23} \Delta q + D_{12} q_{2}$, $A_{1} = D_{12} \sqrt{4+\left( \Delta q \right)^{2}}$,
	$B_{1} = D_{23} \sqrt{1+q_{2}^{2}}$, $A_{2} = D_{12} \sqrt{1+q_{2}^{2}}$, $B_{2} = D_{23} \sqrt{4+\left( \Delta q \right)^{2}}$, $\alpha = \tan^{-1} \left( \frac{\Delta q}{2} \right)$,
	$\beta = \tan^{-1} \left( q_{2} \right)$ and finally, $\Delta q = q_{2} - q_{1} = q_{2} - q_{3}$, $\Delta \omega = \omega_{2} - \omega_{1} = \omega_{2} - \omega_{3}$.
	Note that Eqs.~(\ref{eq:Phase_Diff_Final_1}, \ref{eq:Phase_Diff_Final_2}) represent the  \emph{Adler equation} in two variables~\cite{adler_eqn}. Since, we are interested in finding the behavior of frequency vs coupling, we first assume that oscillators 2 and 3 are phase entrained, that is $\dot{\phi}_{32} \approx 0$, then the beat frequency, $2\pi(\int_{0}^{2\pi}\frac{d\phi_{12}}{\dot{\phi}_{12}})^{-1}$,  between oscillator 1 and 2 is given by:
	
	
	\begin{multline}
		\Omega_{12} = \biggl\{ 
		\left( \Delta \omega(1+\kappa D_{12}) - \sin \phi_{32} D_{23} \right)^{2}\\ 
		\qquad - 4 D_{12}^2 (1+\kappa^2\Delta \omega^2)
		\biggr\}^{\!1/2}
		\label{eq:Frequency_1}
	\end{multline} 
	
	\medskip
	\begin{figure}[h]
		\centering
		\includegraphics[width=0.45\textwidth]{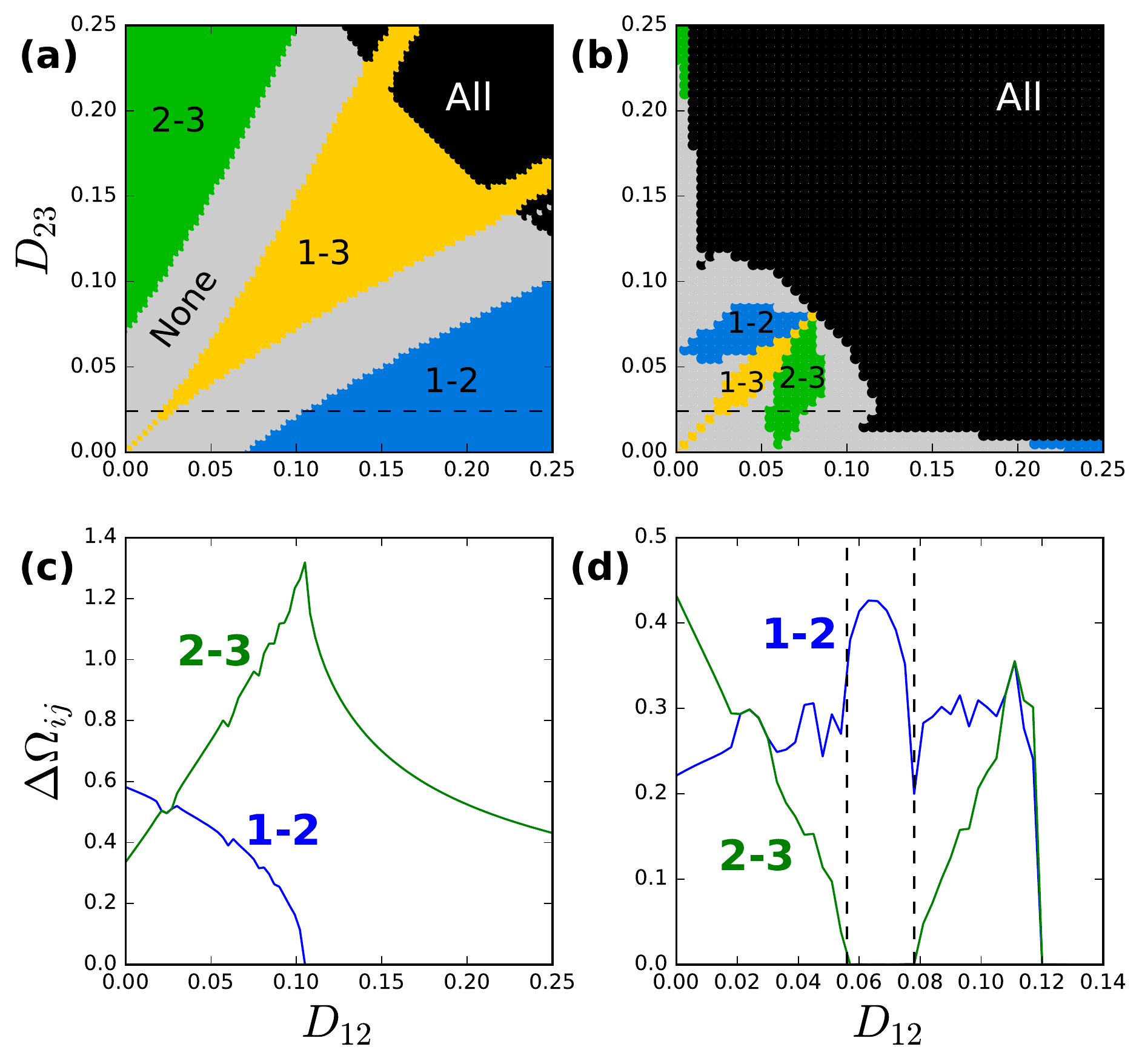}
		\caption{ \small (a,b) : Parameter space plots for Stuart-Landau equations showing, as indicated by the labels(and colors), regions of  $D_{12}$ and $D_{23}$ values for oscillator pairs in phase synchrony or not. On the bottom panel, (c,d) shows the variation of mean relative frequency of pair 1-2 and 2-3, as labeled(and colored), with $D_{12}$ for 
			a fixed value of $D_{23}=0.024$(along horizontal line in the Fig.~\ref{fig:fig3}(a)). Here, $\omega_{1,3} = 1.2$, $\omega_{2} = 0.949$ and $\kappa = -5$(a,c),
			$\kappa=5$(b,d). }
		\label{fig:fig3}
	\end{figure}

	Now, in order to test our hypothesis, \emph{that a non-monotonic dependence of the frequency difference on the coupling strength, arising due to a positive covariance of natural frequency and shear, is responsible for the emergence of weak-winner PS}, we
	define: $q_{j}=\kappa \omega_{j}, ~ j=1,2,3$, where $\kappa=\frac{d q}{d \omega}$ is just a scaling constant. This relation ensures that there is a positive covariance between $\omega$ and $q$ when $\kappa>0$, which is needed for APS to manifest. Substituting, $q_{j}=\kappa \omega_{j}$ and $D_{23}=0.024$ into Eq.~(\ref{eq:Frequency_1}), it can be shown that $\Omega_{12}$ is a non-monotonic function of $D_{12}$ iff $\kappa>0$ (see Appendix A), which is further confirmed by a numerical solution (Fig.~\ref{fig:fig3}(c,d)). Note that the numerically obtained phase diagram of the complex Stuart-Landau system (cf. Fig.~\ref{fig:fig3}(a,b)) represented by Eq.~(\ref{eqn:SL_eqns}) looks very similar to that of the population dynamical system (cf. Fig.~\ref{fig:fig1}(a,c)) represented by Eq.~(\ref{eq:SystemEqns}). For a negative covariance, i.e. $\kappa<0$, we find all the regimes (Fig.~\ref{fig:fig3}(a)) which are also present in Fig.~\ref{fig:fig1}(a) including relay-synchronization. 
	However, for positive covariance, i.e. $\kappa>0$, we obtain quite prominent regions of weak-winner PS (Fig.~\ref{fig:fig3}(b)). This demonstrates clearly, that the presence of APS leads to weak-winner phase synchronization.
	
	

	
	\section{Implication for Networks}
	\label{sec:networks}
	
	As demonstrated earlier, the weak-winner phenomenon is quite generic with respect to the nature of the dynamics of the individual oscillators. However, one might be tempted to think about another aspect of generality which is \emph{topology}. In other words, does this phenomenon hold true for (a) a larger number of oscillators and (b) more complex network topologies? Though the full answer to these questions is beyond the scope of this paper, we present here the first step in this direction by discussing a number of weak-winner patterns that would emerge in larger networks.

\begin{figure}[h]
		\centering
		\includegraphics[width=0.21\textwidth]{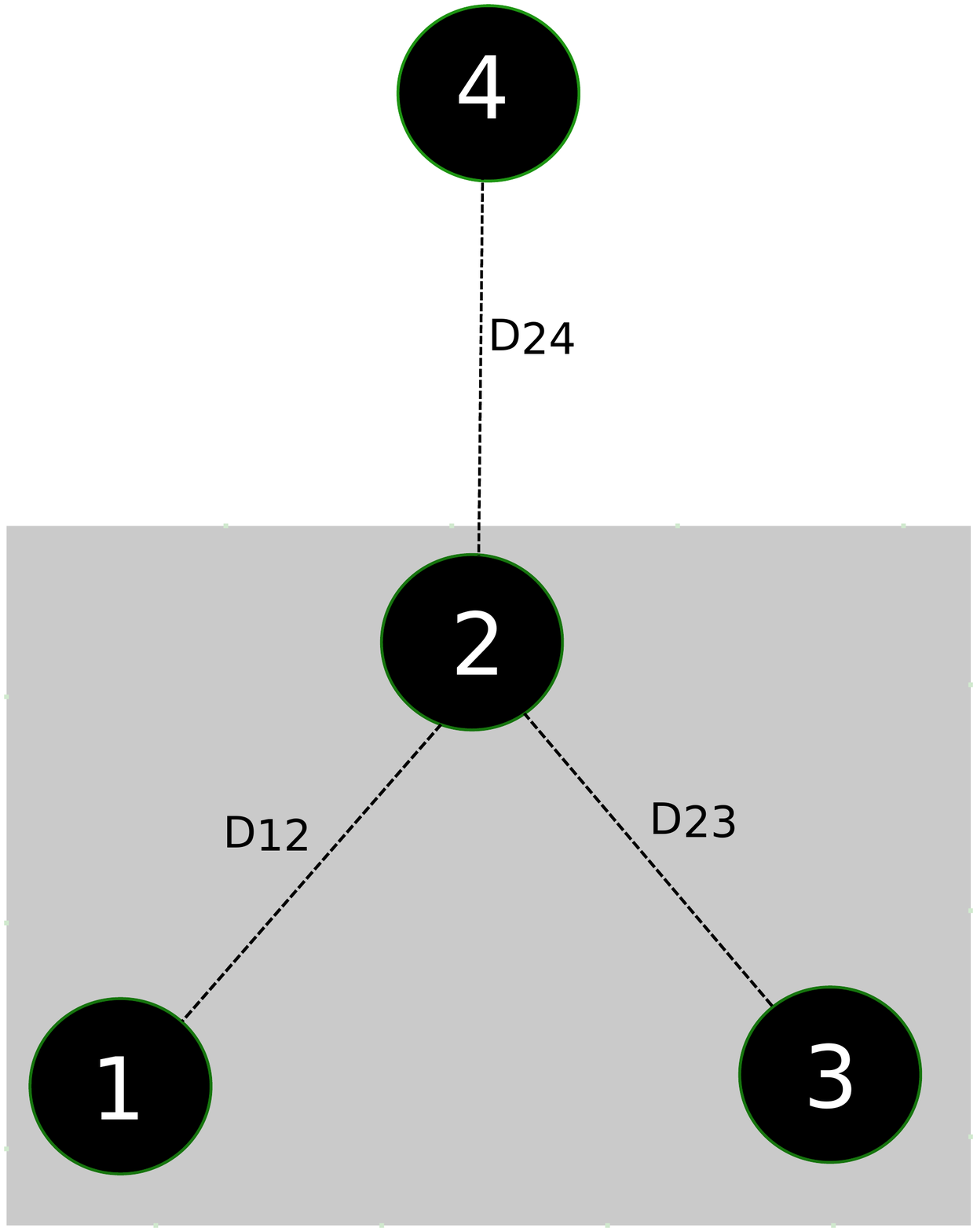}
		\caption{Sketch for realizing the weak-winner phenomenon in complex networks where our original three oscillator system (shaded region) is acting as a network motif and the $4^{th}$ oscillator is acting as a mean-field contribution from the rest of the network. }
		\label{fig:WWN4_schematic}
	\end{figure}

	Specifically, we construct a set-up of four coupled oscillators arranged in such a way that the new set-up can be treated as a combination of our old three oscillator system plus an external fourth oscillator coupled to it, as depicted in the sketch shown in Fig.~\ref{fig:WWN4_schematic}.

\begin{figure}[!h]
		\centering
		\includegraphics[width=0.45\textwidth]{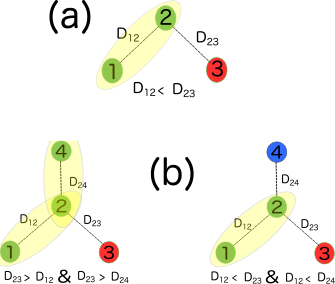}
		\caption{Distinct number of weak-winner phase synchronization patterns that are observed with network of (a) 3 oscillators and (b) 4 oscillators. Nodes of the same color~(shaded in yellow) are phase synchronized while the other nodes are desynchronized. }
		\label{fig:WW4_patterns}
	\end{figure}

	In this set-up, our three oscillator system (shaded region) can serve as a network motif and the $4^{th}$ oscillator encapsulates the mean field contribution of a larger network. To demonstrate the validity of weak-winner phase synchronization in the presence of an external coupling, $D_{24}$ in our case, we simulate the system composed of four Stuart-Landau oscillators arranged in the set-up shown in Fig.~\ref{fig:WWN4_schematic}. For this set-up the observed distinct ``weak-winner" patterns are sketched in Fig.~\ref{fig:WW4_patterns}(b), which shows that for larger networks, not only a pair of weakly linked oscillators can synchronize but also a subset of all weakly linked oscillator pairs can synchronize. 
	
	\begin{figure*}[t]
		\centering
		\includegraphics[scale=0.54]{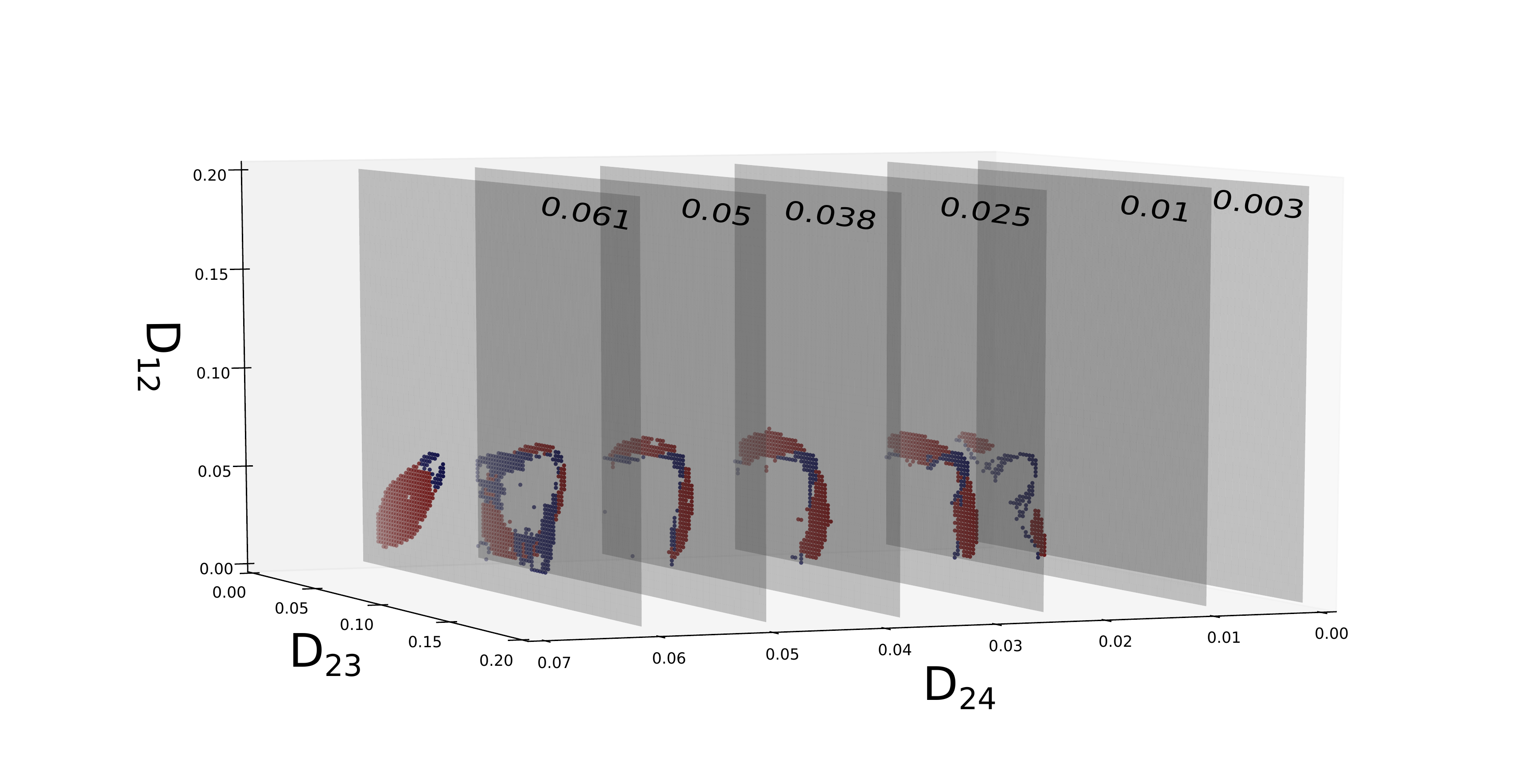}
		
		\caption{Regions of weak winner phase synchronization in a 4-oscillator set-up (as shown in Fig.~\ref{fig:WWN4_schematic}) with our original three oscillators acting as a network motif. As $D_{24}$ is varied, two distinct regions of weak-winner PS~(blue and red colors) are manifested over a significant range of $D_{24}$ values, as shown by parallel planes each corresponding to a fixed value of $D_{24}$.  }
		\label{fig:WW4_regions}
	\end{figure*}
	
	For the four oscillator set-up, regions
	of these distinct weak-winner patterns in the three dimensional coupling space, are presented in Fig.~\ref{fig:WW4_regions}. The blue colored region corresponds to a single pair of oscillators exhibiting weak-winner phase synchronization while the red colored region corresponds to a chain~(of length 3) of oscillators exhibiting weak-winner phase synchronization.

	With even larger networks, the number of ways in which weak-winner phase synchronization can manifest would increase further, giving rise to a wide range of interesting synchronization patterns.

	\section{\label{sec:level4} Discussion and Conclusion}
	The current approach to the phenomenon of phase synchronization in coupled oscillator systems focuses essentially on the interplay of coupling and frequency detuning between the oscillators. However, this approach often overlooks the crucial role played by non-isochronicity~(shear) -- an intrinsic property of an individual oscillator -- in shaping the emergent collective dynamics of the system. For instance, the mechanism behind the emergence of the counter-intuitive weak-winner PS cannot be explained through the current approach. In this context, our study offers not only the underlying mechanism behind weak-winner PS but also sheds some light on the generic question of \emph{how shear influences the phase dynamics?} To the best of our knowledge, the emergence of weak-winner PS is an unprecedented phenomenon, and therefore we anticipate potential applications 
	for a large class of problems where phase synchrony is desirable or in some 
	cases undesirable. For example, in the case of the
	three patches of wildlife equations we used, the coupling
	strengths correspond to the migration rate of predator and
	prey species, and can be interpreted as movement corridors connecting
	different patches of wildlife~\cite{cushman_conservation}. As a conservation strategy, the design of movement corridors should be such that we are able to control the migration intensity of species so that it does not become too low to risk local extinction and not too high either to risk global extinction due to synchronization of populations~\cite{synchrony_dispersal,PS_extinction1,PS_extinction2}.  
	However, the presence of a weak-winner phenomenon could easily make the design of control strategies more difficult as increasing the migration between two patches could induce synchrony among the other patches with weaker migration, which is clearly an undesirable consequence.  
	Although we have shown in Section~\ref{sec:networks} that the 
	emergence of weak-winner in larger complex networks could display several interesting patterns of phase synchrony, we have barely scratched the surface of potential applications/problems that might come up.  One particularly important problem that we could envisage is related to oscillator networks with shear diversity~\cite{shear_diversity}. The heterogeneity in shear and frequency can induce frustration in the oscillatory system and this could result in metastable states (weak-winner like) similar to that of spin-glass systems~\cite{spin_glass1,spin_glass2,spin_glass3,spin_glass4}. 
	
	
	Furthermore, this study is just an initial step in understanding the role of non-isochronicity~(shear)
	in shaping the synchronization behavior of coupled oscillators. For further analysis, we need a better understanding of the
	precise functional relationship between system parameters and the resulting non-isochronicity, i.e., \emph{how shear is determined by the system parameters of an oscillator?}
	Additionally, although having derived {\it sufficient} conditions which help in identifying the regions of the parameter space where one observes weak-winner phase synchronization, it still leaves us with a huge parameter space to explore. To address this in more detail, it would be an interesting and challenging task to derive {\it necessary} conditions
	as well. 
	
	In summary, we disclose a novel and intriguing type of
	phase synchronization in a chain of three coupled oscillators in which
	the weakly coupled oscillators achieve synchrony while the
	strongly coupled ones do not. This result goes beyond the well-known, long established relationship between phase synchronization and coupling strength.
	Further, we have shown that the emergence of
	this unexpected kind of synchronous behavior can be explained 
	in terms of anomalous phase synchronization,
	arising from a complex interplay between shear and natural frequencies
	of the oscillators. The fact that shear and natural frequency are intrinsic properties
	of every oscillator makes the manifestation of the weak-winner phenomenon quite generic. We have validated it by considering oscillators exhibiting different dynamical behaviors such as limit cycle and chaotic dynamics. 
	Some potential applications of weak-winner phase synchronization could include, among
	others, lasers~\cite{mcallister_pre,PS_lasers_roy}, communication systems~\cite{jovic_synchronization}
	and neuronal systems~\cite{neuro_PS1,feudel_chaos}. 
	Lastly, we believe that the mechanism underlying the weak-winner phenomenon would open up a new direction of thinking about the role of non-isochronicity~(shear) as a fundamental feature in shaping the dynamics of any coupled oscillator system.

	\section*{\label{sec:level6}Acknowledgments}
	The authors would like to thank Edward Ott for discussions on a previous version of this manuscript.
	The simulations
	were performed at the HPC Cluster CARL, located at the
	University of Oldenburg (Germany) and funded by the
	Deutsche Forschungsgemeinschaft through its Major Research
	Instrumentation Programs (INST 184/157-1 FUGG) and the
	Ministry of Science and Culture (MWK) of the Lower Saxony
	State. Arindam Saha would like to thank Volkswagen Foundation for the financial support~(Grant
	No. 88459).\\
	\bigskip
	\section*{\label{sec:level7}Appendix A:~Behavior of the frequency difference ($\Omega_{12}$) as a function of $D_{12}$}
	In this section, we demonstrate the existence of Anomalous Phase Synchronization (APS) in our model system (Eq.~\ref{eqn:SL_eqns}) by establishing the non-monotonic behavior of $\Omega_{12}$ as a function of $D_{12}$. We start by expanding Eq.~\ref{eq:Frequency_1} as :
	\begin{widetext}
		\begin{equation}\label{sup_eq:Frequency_2}
		\Omega_{12} = \sqrt{\left[ 2\kappa \Delta \omega (\Delta \omega - D_{23}\sin\phi_{32})\right]D_{12} -4D_{12}^{2} + \left[ \Delta \omega^2 - 2\Delta \omega \sin\phi_{32}D_{23} + (\sin\phi_{32}D_{23})^2 \right]}
		\end{equation}
	\end{widetext}
	In deriving Eq.~\ref{eq:Frequency_1}, we assumed  $\dot{\phi}_{32} \approx 0$, which holds true for an interval, $[D_{12}^1, D_{12}^2]$(see Fig.~3(d)). Therefore, for this interval, $\sin\phi_{32}$ becomes a constant. Also, since we are interested in finding the behavior of $\Omega_{12}$ as $D_{12}$ is varied, we keep $D_{23}$ fixed at a value of $0.024$ (as shown by the horizontal line in Fig.~3(d)). Rearranging all the constant terms yields:
	\begin{equation}
	\Omega_{12} = \sqrt{-4D_{12}^2 + \mu D_{12} + \nu}
	\label{sup_eq:Frequency_3}
	\end{equation}
	where, $\mu = 2\kappa(0.063-0.006\sin\phi_{32})$ and $\nu = (0.063-0.012\sin\phi_{32}) + 0.0006\sin^2\phi_{32}$. Since the function of the relative frequency is \emph{quadratic} in coupling strength, we can check if an extremum exists in the interval ($[D_{12}^1, D_{12}^2]$), which would confirm the presence of non-monotonicity. Therefore, the problem now reduces to finding the extremum of $g(D_{12}) = -4D_{12}^2 + \mu D_{12} + \nu $, which gives us $D_{12}= \frac{\mu}{8}$. Observe, that the value of $D_{12}$ is positive(valid solution) if and only if $\mu>0$ which holds true when $\kappa >0$ and this validates our claim that APS exists only for $\kappa>0$.
	\section*{\label{sec:level8} Appendix B:~Coupled Van der Pol oscillators}
	Here, we present the phase diagram of three coupled Van der Pol oscillators exhibiting limit cycle oscillations. The coupling set-up used here is the same as in the model of chaotic oscillators given by Eq.(1) in the main text. The governing equations are thus represented by,
	\begin{subequations}\label{sup_eq:vdp}
		\begin{align}
			\dot{x}_{i} &= y_{i} + \sum_{i=1}^{3}D_{ij}(x_{j}-x_{i}),\\
			\dot{y}_{i} &=  a_{i}(1-x_{i}^{2})y_{i} - b_{i}^{2}x_{i} +\sum_{i=1}^{3}D_{ij}(y_{j} -y_{i} )
		\end{align}
	\end{subequations}
	\begin{figure}[h]
		\centering
		\includegraphics[width=0.42\textwidth]{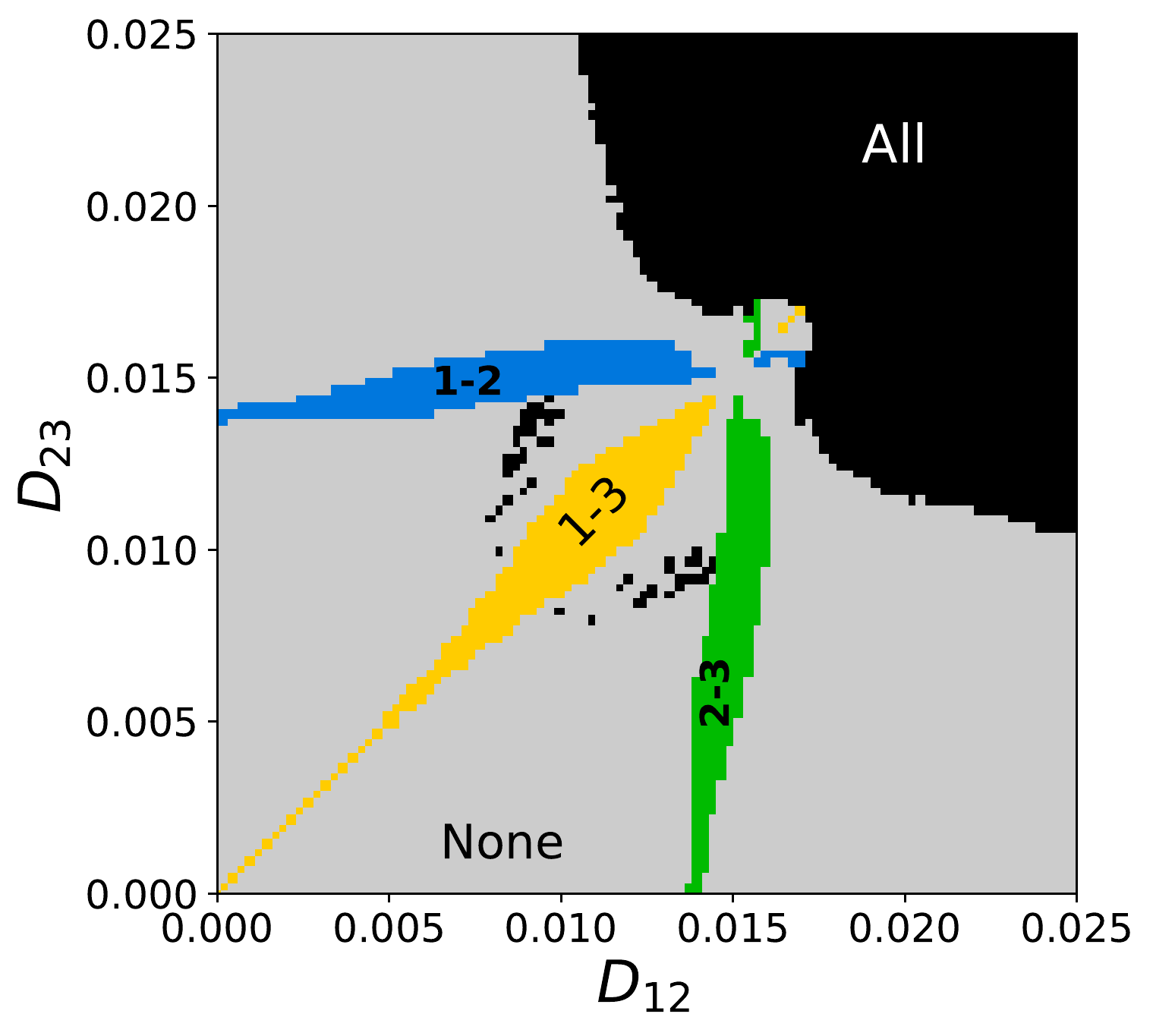}
		\caption{ \small Parameter space plots showing, as indicated by the labels (and colors), regions of  $D_{12}$ and 
			$D_{23}$ values for oscillator pairs in phase synchrony or not. Parameters are  {\bf a} =(1,1,1) and ${\bf b}=(1.0,0.92,1.0)$.  }
		\label{fig:fig4}
	\end{figure}
	To observe weak-winner phenomenon, we set $a_{i} = \kappa b_{i}$ with $\kappa = 4.0 ~\forall i=1,2,3$ and ${\bf{b}} = \left[1,0.92,1\right]$. Islands of weak-winner solutions~(green and blue regions) can be clearly seen in Fig.~\ref{fig:fig4}.
	\bigskip

	\section*{\label{sec:level8} Appendix C:~Intuition behind Anomalous Phase Synchronization}
	The aim of this section is to promote an intuitive understanding of anomalous phase synchronization (APS). First we introduce the concept of \emph{isochron (shear)} which is essentially the dependence of rotation speed on amplitude. Formally, \emph{isochrons} are defined as a family of curves in phase space where all points on each curve represent a unique phase~\cite{winfree_shear}.
	To demonstrate this we take the Stuart-Landau oscillator given by, 
	\begin{equation}\label{s_eq:SL_eqn}
	\dot{z} = z\left[1+i(\omega+q) -(1+iq)|z|^2  \right]
	\end{equation}
	which in polar coordinates becomes,
	\begin{align}
		\dot{r} &= r( 1-r^{2}),\\
		\dot{\theta} &= \omega + q \left( 1-r^{2} \right) .
	\end{align}
	\begin{figure}[ht]
		\centering
		\includegraphics[width=0.49\textwidth]{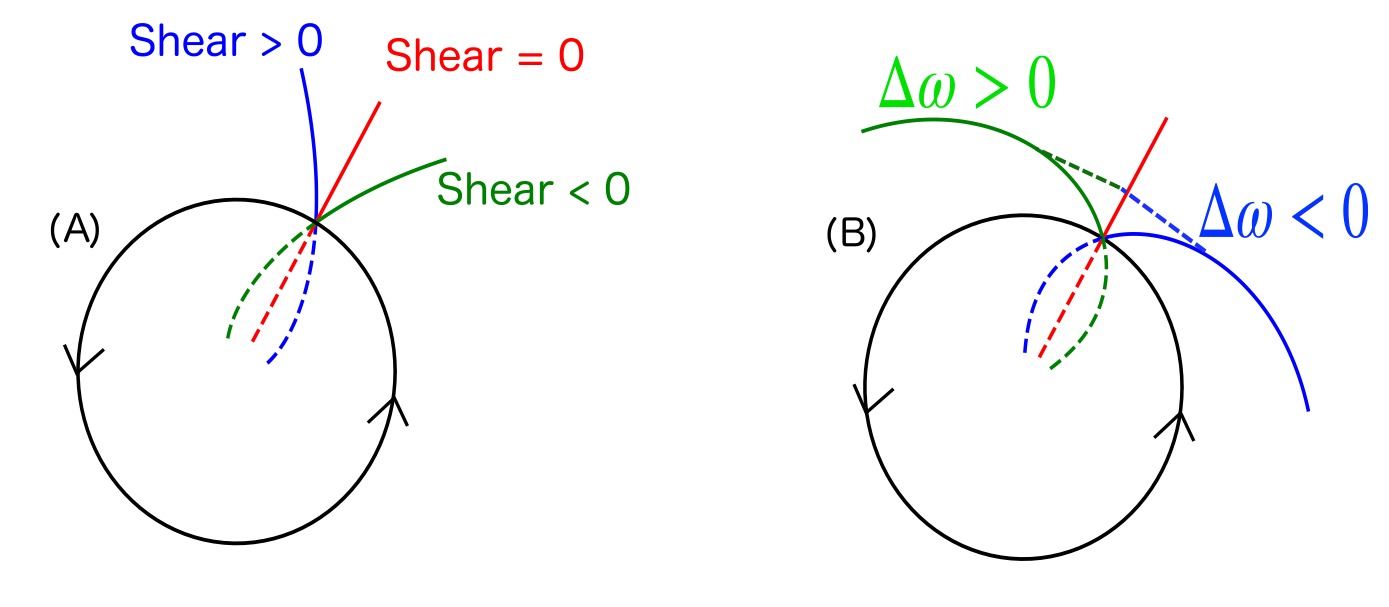}
		\caption{ \small (A) Isochrons (B) frequency variation with amplitude $\omega(r)$, of a Stuart-Landau oscillator (Eq.~\ref{s_eq:SL_eqn}) for positive (blue), negative (green) and zero shear (red) as drawn on polar coordinates ($r,\theta$). The black curve represents the limit cycle solution of the oscillator with $r=1$.}
		\label{fig:fig5}
		
	\end{figure}

	This oscillator has a stable limit cycle solution at $r=1$.
	The phase can be defined in the neighborhood of the limit cycle attractor as~\cite{sync_thesis}: $\phi =\theta - q\ln{r}$, which on the limit cycle becomes just `$\theta$' as $r=1$. Therefore, a typical \emph{isochron} representing a constant phase ($\phi^*$) is described as: 
	\begin{equation}\label{sup_eq:isochron}
	I_{\phi^*} = \phi^* = \theta - q\ln{r} 
	\end{equation}

	When $q=0$, the isochron has no radial component which means that rotation speed is independent of the position in the neighborhood of the limit cycle. In Fig.~\ref{fig:fig5}(A), we show how isochrons change as shear is introduced in the system. For positive shear ($q>0$), the oscillator's instantaneous frequency increases (decreases) as we move radially inwards (outwards) from the limit cycle. This change in frequency is captured by $\Delta\omega=q(1-r^2)$ as shown in the Fig.~\ref{fig:fig5}(B). However, for negative shear ($q<0$), the system gets slower (faster) as it moves inwards (outwards) from the limit cycle. Note that, fast and slow are always relative to the case of zero shear ($q=0$), where the frequency is independent of amplitude.

	Now, we extend this picture to two diffusively coupled Stuart-Landau oscillators having frequencies $\omega_{1}$ and $\omega_{2}$ respectively with $\omega_{1}<\omega_{2}$. Further,
	we set $q_{i}=\kappa \omega_{i}$ for $i=1,2$, so that $\Cov[q, \omega]$ can be positive, negative or zero depending on the values of $\kappa$.
	As before, the relative change in the frequency of the oscillator due to shear is measured by $\Delta\omega_{i}(r_{i})$. However, in this setup, shear for an oscillator is not just a constant but depends on its \emph{natural frequency} $\omega$. To illustrate the effect of shear on the resulting behavior of the coupled system, we consider the three following cases: 
	
	\begin{figure}[ht]
		\centering
		\includegraphics[width=0.49\textwidth]{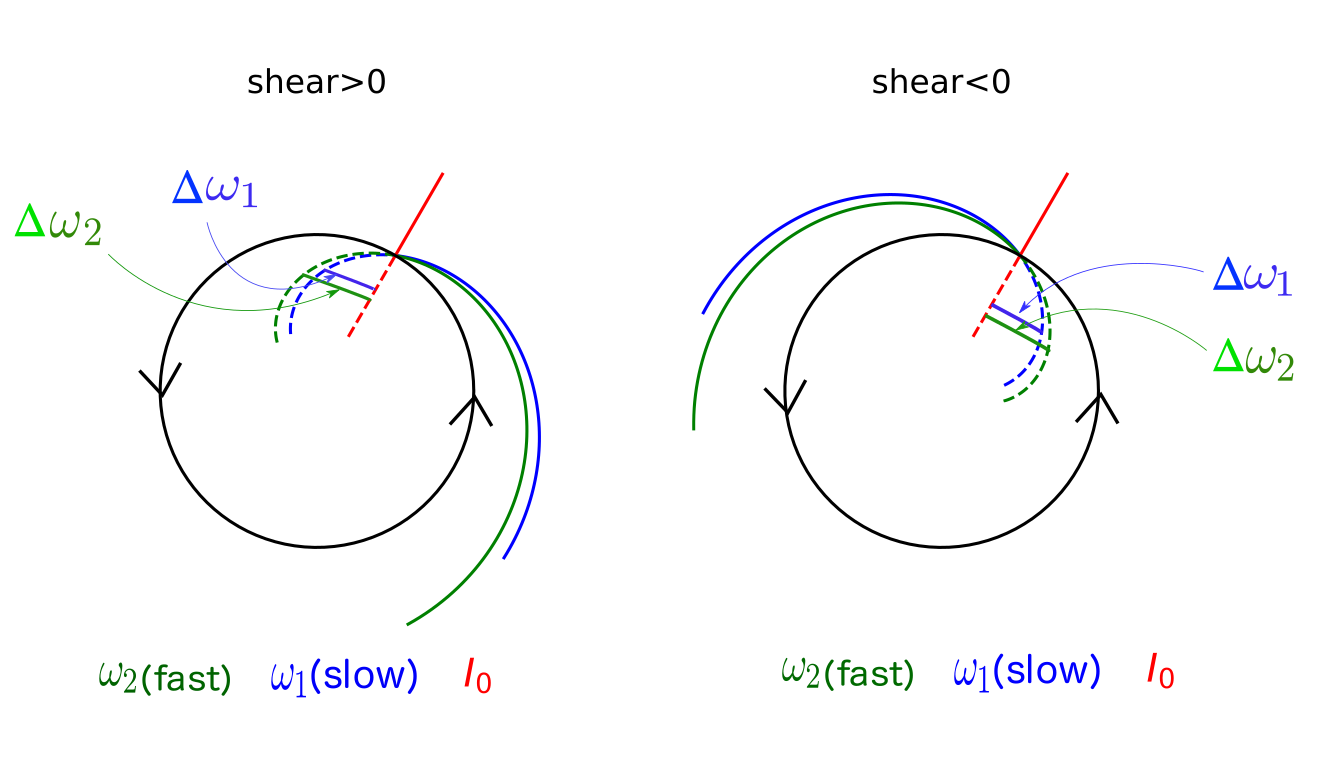}
		\caption{ \small Frequency variation with amplitude for two diffusively coupled Stuart-Landau oscillators for positive ($\kappa>0$) and negative ($\kappa<0$) shear as drawn on polar coordinates ($r,\theta$). The instantaneous frequency curves corresponding to fast and slow oscillators are represented by green and blue colors respectively. The red one represents the shearless case. The black curve represents the limit cycle solution with $r=1$ for the uncoupled oscillator.}
		\label{fig:fig6}
	\end{figure}

	\begin{enumerate}
		\item ~$ \bf  \kappa=0~(shearless)$:\\\\
		This is a trivial case where both oscillators do not experience any shear. Here the diffusive coupling would have a trivial impact on the dynamics, i.e. \emph{increasing the coupling would slow down the fast oscillator and speed up the slow one until both oscillators lock to a common frequency and start oscillating synchronously}.
		\medskip
		
		\item ~$ \bf  \kappa>0~(positive~shear)$ :\\\\
		In this case, both oscillators experience different shear ($q_{1,2} = \kappa \omega_{1,2}$). Since $\omega_{1}<\omega_{2}$, we have $q_{1}<q_{2}$. Frequency variation for both oscillators are shown in Fig.~\ref{fig:fig6} marked by their respective colors. For positive shear, both oscillators speed up but by different amounts ($\Delta\omega_{1}<\Delta\omega_{2}$) when they move inwards, away from the limit cycle. It is worth noting here that due to coupling the 
		oscillators are always pushed inwards as shown by the variation of mean amplitude as a function of coupling strength (Fig.~\ref{fig:fig7}).
		
		Therefore, for lower coupling strengths the oscillators are almost always inside the limit cycle and then the shear comes into play which in this case \emph{widens} their initial frequency difference. Eventually, for high enough coupling, the oscillators are pulled back to follow the limit cycle where the effect of shear vanishes and they manage to synchronize. This is essentially the mechanism behind APS.
		\bigskip
		\item ~$\bf  \kappa<0~(negative~shear)$:\\\\
		The explanation for the case of negative shear is quite similar to that of positive shear except here the frequency variation (Fig.~\ref{fig:fig6}(right)) is such that both the oscillators slow down with faster one slowing down by a larger amount than the slow one.
		\begin{figure}[h]
			\centering
			\includegraphics[scale=0.42]{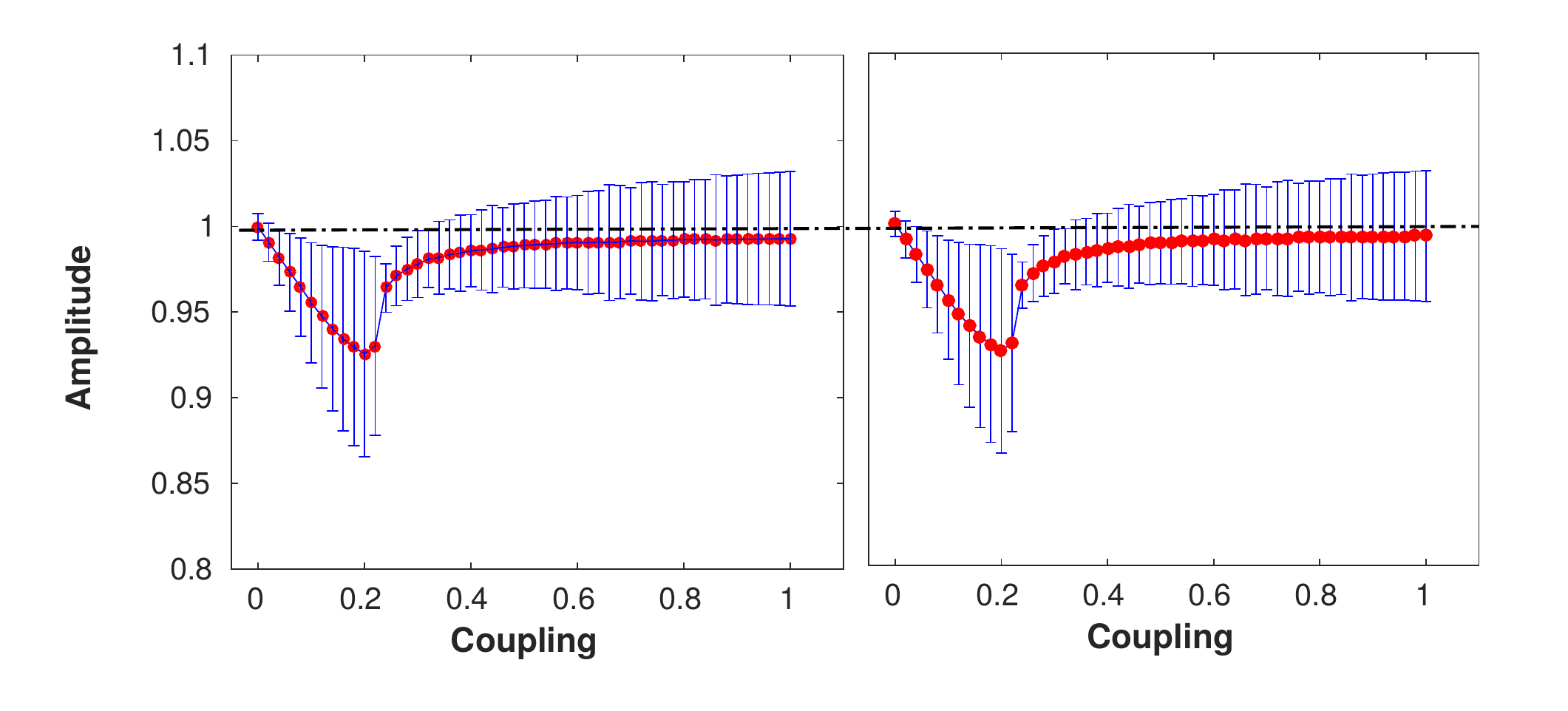}
			\caption{ \small Variation of amplitudes ($r1$ - left, $r2$ - right) with coupling strength for two diffusively coupled Stuart-Landau oscillators. The natural frequencies of the oscillators are: [0.95, 1.2] and $\kappa=5$. Red points indicate mean values and error bars ($blue$) represent standard deviation.}
			\label{fig:fig7}
		\end{figure} 
		Therefore, by contrast to the previous case of positive shear, the weak coupling would \emph{shrink} their initial frequency difference. This means that negative shear synchronizes the system at a coupling strength even lower than that of zero shear.
	\end{enumerate}
	
	\section*{\label{sec:level9} Appendix D:~Connection to Short wavelength Bifurcation}
	One might suspect that weak-winner phase synchronization arises as a consequence of a short wavelength bifurcation~(SWB) of the type discovered for diffusively coupled oscillators~\cite{Sync_MSF}. Briefly, SWB is a bifurcation of the synchronized dynamics residing on the invariant synchronization manifold in a system of coupled identical oscillators, where upon increasing coupling strength, the systems loses synchrony. It occurs when the eigenvalue corresponding to the smallest spatial Fourier mode becomes positive. As a consequence of this bifurcation, the system can only be synchronized for a bounded range of coupling strengths. 
	
	\medskip
	At first, this could suggest an explanation of weak-winner PS due to the fact that in a system with mixed coupling strengths, the weaker coupling is still in the intermediate range~(where synchrony is possible) while the stronger coupling is already beyond that range. However, via numerical simulations we demonstrate below that the pairs of coupled limit cycle and chaotic oscillators never lose synchronization after its onset, within the investigated range of coupling strengths. 
	
	\medskip
	\begin{enumerate}
		\item \underline{Limit cycle case}:
		\begin{equation}\label{eq:SL_2osc}
		\dot{z}_{j} = z_{j}\left[1+i(\omega_{j}+q_{j}) -(1+iq_{j})|z_{j}|^2  \right] + D(z_{k} -z_{j} )
		\end{equation}
		\item \underline{Chaotic food web case}:
		\begin{subequations}\label{eq:chaotic_2osc}
			\begin{align}
				\dot{x}_{j} &= a_{j}x_{j} - \frac{\epsilon_{1}x_{j}y_{j} }{(1+k_{1}x_{j} )}\\
				\dot{y}_{j} &= -b_{j}y_{j} + \frac{\epsilon_{1}x_{j}y_{j} }{(1+k_{1}x_{j} )} -\epsilon_{2}y_{j}z_{j} + D(y_{k} -y_{j} )\\
				\dot{z}_{j} &= -c_{j}(z_{j}-\zeta_{j}) + \epsilon_{2}y_{j}z_{j} +D(z_{k} -z_{j} ) 
			\end{align}
		\end{subequations}
	\end{enumerate}
	\medskip
	where, $j,k=1,2$~($j\ne k$) and D is the coupling strength. The system parameters for the limit cycle case are: $\omega_{1} = 0.949$, $\omega_{2} = 1.2$, $q_{1,2}=\kappa\omega_{1,2}$ with $\kappa=4.0$ and for the chaotic oscillator case: $a_{1}=a_{2}=1.0$, $b_{1}=0.9$, $b_{2}=1.3$, $c_{1}=c_{2}=10.0$, $\epsilon_{1}=0.25$, $\epsilon_{2}=1.0$, $\zeta_{1}=\zeta_{2}=0.006$. 
	We would like to emphasize here that the system parameters of both oscillators are such that there is a slight detuning in the natural frequencies of the oscillators which essentially means they are non-identical in contrast to the case considered by Pecora and Carroll~(1998) in their master stability approach~\cite{Sync_MSF}.
	
	For both the systems~(Eq.~16, 17), we observe the following quantities as the coupling between the oscillators is varied:
	\begin{enumerate}
		\item Average frequency difference: $\langle f_{1} - f_{2} \rangle_{t}$, where $f_{1}$ and $f_{2}$ represent the instantaneous frequencies of oscillator 1 and 2, respectively. The average of the frequency difference is taken over time ``t steps" after transients have settled. When this quantity approaches zero, we have phase synchronization between oscillators.
		\item Root mean squared deviation: This quantity measures the extent of complete synchronization in the system. Mathematically, it is given by:
		\begin{equation*}
			Z_{sync} = \sqrt{\frac{1}{NT}\sum_{t=1}^{T}\sum_{i=1}^{N}\left\Vert({\bf X_{i}(t)}-\langle {\bf X(t)}\rangle_{i})\right\Vert^2}
		\end{equation*}
		where, $\langle . \rangle_{i}$ is the average over the number of oscillators N, $\left\Vert . \right\Vert$ is the euclidean norm and $1<t<T$ is the time interval after transients have settled.
		In case of complete synchronization, i.e.  $\bf X_{1} = X_{2} = ... = X_{N}$ $\forall$ t, $Z_{sync} \to 0$ asymptotically.
	\end{enumerate}
	\medskip
	\begin{figure}[h]
		\centering
		\includegraphics[width=0.55\linewidth]{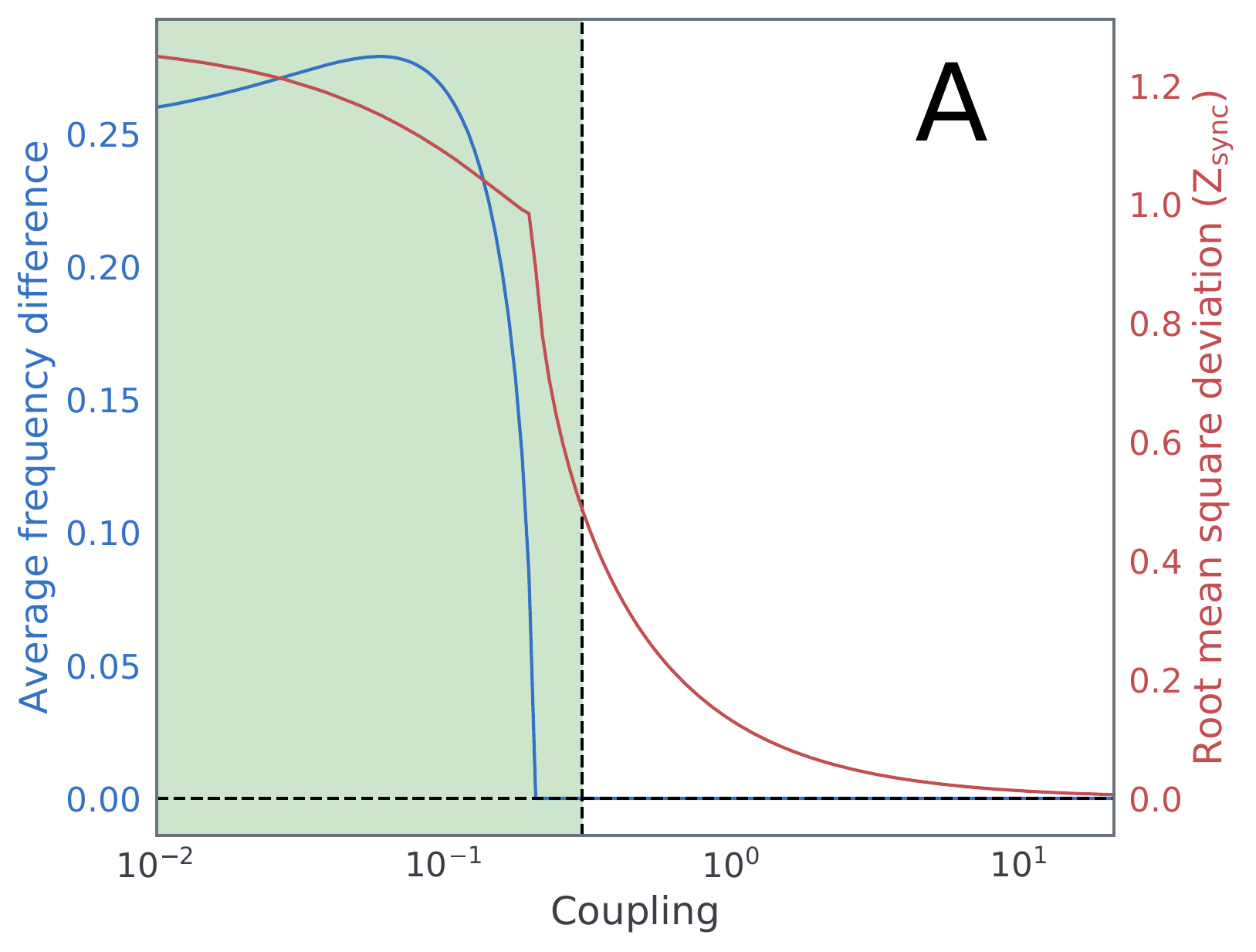}
		
		\includegraphics[width=0.55\linewidth]{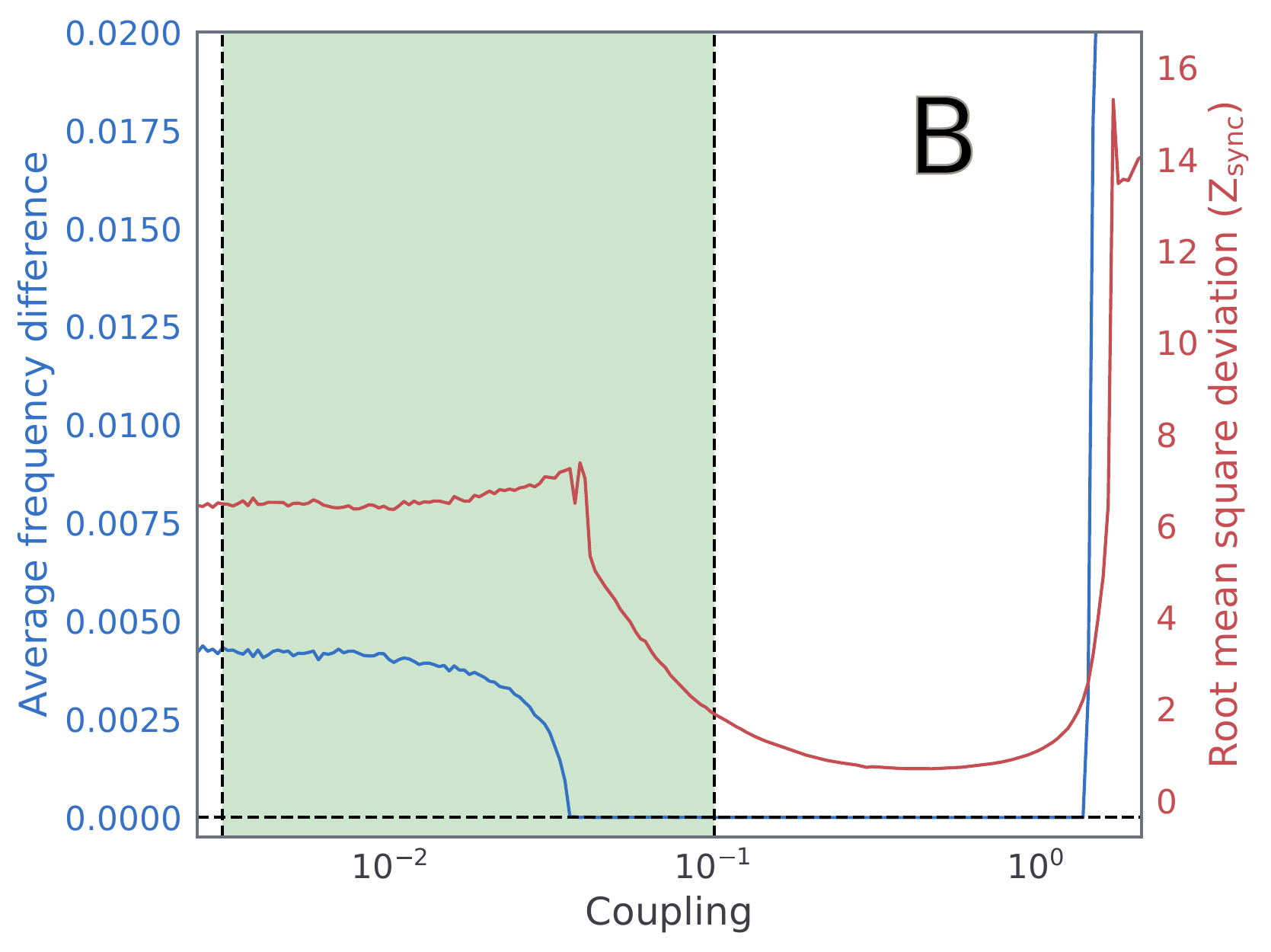}
		\caption{\small Variation of the average frequency difference (blue curve) and the root mean squared deviation~($Z_{sync}$, red curve) with the coupling strength for two diffusively coupled oscillators. (A) coupled limit cycle oscillators (Eqn.~\ref{eq:SL_2osc}) (B) coupled chaotic food web model (Eqn.~\ref{eq:chaotic_2osc}). The green shaded area depicts 
			the range of coupling strength considered in our study.
			\label{fig:reply_1}  
		}
	\end{figure}
	
	\medskip
	The simulation results are presented in Figure~\ref{fig:reply_1} and there are a couple of things to be noted here:
	
	\medskip
	\begin{itemize}
		\item For the complete range of coupling strengths used in our study~(green shaded region in Fig.~\ref{fig:reply_1}), the system composed of two oscillators never lose phase synchrony~(indicated by the blue curve) once established, as the coupling is increased. This clearly rules out the involvement of SWB for both, limit cycle as well as for chaotic oscillators.
		
		\item For coupling strengths beyond our studied window, we anticipate that, as coupling is increased the oscillator's amplitude will also tend to synchronize which is reflected by a decrease in $Z_{sync}$. In case of the limit cycle system~(Fig.\ref{fig:reply_1}-A), upon increasing coupling further, the system does not lose its synchrony. However, for the chaotic coupled oscillator case~(Fig.~\ref{fig:reply_1}-B), the system undergoes a loss of synchrony upon increasing coupling strength which might be a signature of SWB. Please note that this happens at coupling values much higher than the ones used in our study.
		
		\medskip
		
		\item  Moreover, while the SWB reported by Pecora and Caroll~\cite{Sync_MSF} was found for identical oscillators and identical coupling strengths, the weak-winner PS that we present here, emerges only when oscillators are detuned sufficiently and the coupling strengths are not identical.
	\end{itemize}
	
	
	%


	
	%

\end{document}